\documentclass[superscriptaddress,showpacs,aps,twocolumn,prb,floatfix,nobibnotes]{revtex4-2}
\usepackage{amssymb}
\usepackage{amsmath}
\usepackage{mathtools}
\usepackage{graphicx}
\usepackage{bm}
\usepackage{fontenc}
\usepackage{color}
\usepackage{ulem} 
\usepackage{varioref}
\usepackage{xypic}
\usepackage[hidelinks]{hyperref}
\newcommand\RR{\mathbb R}

\begin{document}
\title{Exact analytical edge states in the extended Su–Schrieffer–Heeger model}

\author{P. A. Grizzi}
\affiliation{Centro At\'{o}mico Bariloche, C.N.E.A, Bariloche, Rio Negro and Consejo Nacional de Investigaciones Cient\'{\i}ficas y T\'ecnicas, Argentina}

\author{A. A. Aligia}
\affiliation{Centro At\'{o}mico Bariloche, C.N.E.A, Bariloche, Rio Negro and Consejo Nacional de Investigaciones Cient\'{\i}ficas y T\'ecnicas, Argentina}
\affiliation{Instituto de Nanociencia y Nanotecnolog\'{\i}a
CNEA-CONICET, GAIDI,
Centro At\'{o}mico Bariloche, 8400 Bariloche, Argentina}

\author{P. Roura-Bas}
\email{pabloroura@integra.cnea.gob.ar}
\affiliation{Centro At\'{o}mico Bariloche, C.N.E.A, Bariloche, Rio Negro and Consejo Nacional de Investigaciones Cient\'{\i}ficas y T\'ecnicas, Argentina}

\begin{abstract}
We investigate the topology of the different phases of the extended Su–Schrieffer–Heeger (eSSH) model, which includes hopping processes between translationally inequivalent atoms beyond nearest neighbors.
Exact analytical expressions for the edge states of a semi-infinite eSSH chain are derived, with wave functions that decay exponentially from the boundary with a unit-cell decay factor $z$.
From the winding number of the bulk Hamiltonian under periodic boundary conditions, we determine the topological phase diagram and establish the bulk–boundary correspondence: changes in the winding number coincide with bulk gap closings and with the condition $|z|=1$ for the edge-state solutions.
For finite chains, we further obtain analytical, approximate expressions for the low-energy edge states, which are shown to be highly accurate.
\end{abstract}


\maketitle

\section{Introduction}\label{section-intro}
\noindent

Topological materials have attracted a lot of interest in the last years
\cite{Qi,ten-fold-way,shen,Ando}.
Experimentally, topological states of matter have been investigated using
for instance photonic lattices \cite{Recht,Lu,has,li,li2},
scanning tunneling microscopy \cite{mis,Zhao,eden} and cold atoms \cite{Atala,loh,naka,vie}.

A topological insulator (TI) is a gapped phase of matter characterized by a nontrivial bulk topological invariant. For noninteracting systems (as in the rest of this manuscript), a finite TI hosts symmetry-protected gapless edge states \cite{ten-fold-way}. In interacting systems, such as the interacting Su–Schrieffer–Heeger (SSH) model, the charge gap need not close, but zero-energy spin excitations can persist at the edges.\cite{manma,aligia-2}

These edge states, described by extended wave functions, are \textit{protected} in the sense that their presence is guaranteed against possible deformations of the Hamiltonian, as long as the symmetries of the Hamiltonian are preserved. Although the definition of a TI is given in terms of finite systems, described by finite-size Hamiltonians, there exists a correspondence with the bulk Hamiltonian---the translation-invariant, momentum-space representation of the infinite system---through the bulk--boundary correspondence (BBC) \cite{hatsugai,Ryu,kane}. The BBC states that topological invariants constructed from the bulk Hamiltonian---integers that are stable against Hamiltonian deformations in the aforementioned sense---determine the existence and number of protected boundary states in a finite system. Thus, the insulator is called \textit{topological} in the sense that its bulk Hamiltonian is characterized by some topological invariant.

Among the simplest systems exhibiting topological phases is the SSH model, which describes a one-dimensional dimerized lattice of spinless fermions with alternating nearest-neighbor couplings \cite{ssh}.
The BBC has been extensively studied in
the SSH model by comparing the associated topological invariant—the
winding number $\delta$—of the bulk model with the number of zero-energy states—the
topologically relevant ones—in finite chains \cite{asboth}.

\begin{figure}[htb]
\begin{center}
\includegraphics[width=1.15\columnwidth]{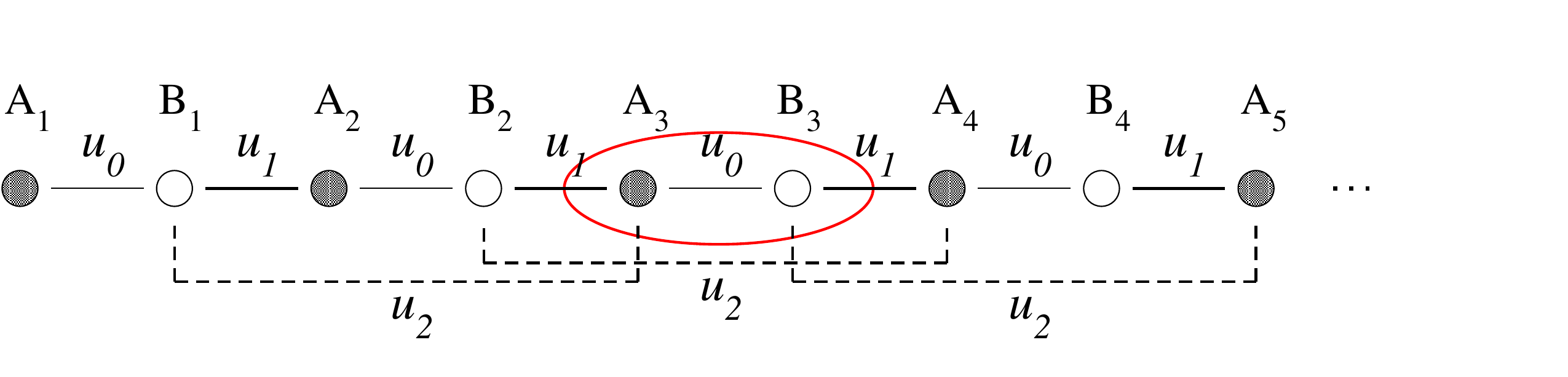}
\end{center}
\caption{Color online. Cartoon of the semi-infinite chain with the hopping amplitudes between the different sites and cells. The red solid ellipse indicates the choice of the unit cell from which the Bloch Hamiltonian is defined. }
\label{finite-chain-cartoon}
\end{figure}

Recently, several works have investigated generalizations of the Su–Schrieffer–Heeger (SSH) model that include hopping processes between atoms that are not equivalent under lattice translations in a periodic system, extending beyond nearest-neighbor distances \cite{li,li2,maffei,ghosh,malakar,joshi,chang,he}. Experimental realizations of various generalized SSH models were achieved using three-dimensional femtosecond laser direct writing by C.~Li and collaborators, as reported in Ref.~\cite{li}. In particular, the so-called extended Su–Schrieffer–Heeger (eSSH) Hamiltonian, which incorporates couplings beyond nearest neighbors, was shown to accurately model the experiments.
The version of the model including next-nearest-neighbor coupling between unit cells is sketched in Fig.~\ref{finite-chain-cartoon}.

The authors validated the model by demonstrating agreement between the expected pair of topologically protected edge states in a finite chain and those observed experimentally.
The edge states are inferred from the spectrum of a $32$-site chain, corresponding to the size of the experimental setup. Furthermore, the authors tuned the coupling intensities in the experimental setup to match the number of edge-state pairs associated with the largest long-range coupling.

The work of S. Li \textit{et al.}, \cite{li2}, studies tunable atomic ensembles to generate one-dimensional bipartite superradiance lattices. They show that several various types of the eSSH model can be realized. In particular, they report a topological phase transition from a topological trivial regime with winding number
$\delta=0$ to a topological non-trivial regime with $\delta=2$.

The same model, as well as a variant with four atoms per unit cell, was studied by Joshi and Nag \cite{joshi}. A related model including second-nearest-neighbor hopping and Coulomb repulsion has been investigated more recently \cite{chang}. In addition, the noninteracting case with hopping terms extending up to fourth-nearest neighbors has been employed in a study of quantum metrology \cite{he}.

In this work, we derive exact analytical expressions for the edge states of the
eSSH model. To this end, we employ the method introduced by Alase \textit{et al.} \cite{alase-1,alase-2}, which allows one to obtain closed-form expressions for the zero-energy eigenstates—the topologically relevant states—of a semi-infinite chain described by the eSSH Hamiltonian. The corresponding wave functions decay exponentially away from the edge, with a factor $z$ for each additional unit cell.

From the winding number of the bulk Hamiltonian under periodic boundary conditions (PBC), we determine the topological phase diagram and discuss the bulk–boundary correspondence. In particular, changes in the winding number occur at parameter values for which the bulk gap closes under PBC, and simultaneously the decay factor of the semi-infinite-chain solution satisfies
$|z|=1$.

Furthermore, from the analytical structure of the edge states, we are able to explain several notable features observed in the work of C.~Li \textit{et al}. reported in Ref.~\cite{li}. In particular, for finite chains we derive analytical, approximate—yet highly accurate—expressions for the low-energy edge states and their energies, which show excellent agreement with the numerical results of Ref.~\cite{li}. In addition, we analyze the edge states and the experimentally observed topological phase transition reported by S.~Li and co-workers in Ref.~\cite{li2}.

The paper is organized as follows. In Section~\ref{section-extended-ssh}, we discuss general aspects of the extended Su–Schrieffer–Heeger model and its topologically distinct phases. In Section~\ref{section-edge-states}, we derive the topological edge states of a semi-infinite chain associated with the bulk eSSH model. In Section~\ref{section-results}, we apply this approach to two different experimental realizations of the model. Finally, Section~\ref{section-summary} summarizes the main results of this work.

\section{Extended Su–Schrieffer–Heeger (eSSH) model and its bulk winding number $\delta$}\label{section-extended-ssh}
One-dimensional two-band chiral models with winding number greater than one can be constructed by introducing long-range hopping processes beyond the simplest nearest-neighbor one.  The simplest example is a generalization of the SSH model that includes hopping to the $j$-neighbor between atoms that are not equivalent under translational symmetry, with amplitudes $u_j$ with $j\in\{1,\dots,M\}$, in addition to the intracell coupling $u_0$. The hopping amplitudes are assumed to be real, ensuring that the system preserves time-reversal symmetry, which, as discussed below, is crucial for determining the topological class of the model.

Note that with the choice of unit cell (AB instead of BA in Fig. \ref{finite-chain-cartoon}), the hoppings $u_j$ with $j \geq 2$ connect the second site of the unit cell only with the first site of a unit cell at the right,
while the first site is connected only with the second site of a unit cell at the left.
This ordering is inverted with the alternative choice of unit cell.

Following Ref. \cite{li} and the choice of unit cell shown in Fig.~\ref{finite-chain-cartoon}, the Bloch Hamiltonian matrix takes the form:
\begin{eqnarray}\label{chiral-bloch-ham}
 {H}(k)=\left(\begin{array}{cc}
        0 & Q(k)\\
        \overline{Q(k)} &0
      \end{array}\right),
\end{eqnarray}
where $Q(k)=d^1(k)-id^2(k)=d(k)\mbox{exp}(-i\phi(k))$, $d(k)=|Q(k)|$ and vanishes only at phase transitions, and $\overline{Q(k)}$ is the complex conjugate of $Q(k)$. The real functions $d^1(k)$ and $d^2(k)$ are given by:
\begin{equation}\label{lr-ssh-equations}
\begin{array}{l}
d^1(k) = \sum_{j=0}^{M} u_j\mbox{cos}(jk) \\[8pt]
d^2(k) = \sum_{j=0}^{M} u_j\mbox{sin}(jk).
\end{array}
\end{equation}

The model possesses both time-reversal and particle–hole symmetries, with the corresponding symmetry operators squaring to the identity. It also exhibits the composite chiral symmetry (see Sec.~III of Ref.~\cite{malakar} and Ref.~\cite{avron}). As a consequence, within the ten-fold way classification, the model belongs to class BDI of topological insulators, characterized by an integer $\mathbb{Z}$ topological invariant~\cite{ten-fold-way}.

It is worth mentioning that, while the ten-fold way classification formally relies on the assumption of a sufficiently large number of bands, the winding number remains the appropriate topological invariant even in the case of chiral two-band models. Therefore, the homotopy classes of Bloch Hamiltonians in Eq.~(\ref{chiral-bloch-ham}) can be labeled by their winding number. For further details, see the work of Avron and Turner in Ref.~\cite{avron}. 

Furthermore, the model exhibits an additional discrete symmetry: inversion through the center of each bond.

The topological invariant---the winding number---is given by:
\begin{equation}\label{delta}
\delta = \frac{1}{2\pi} \int_0^{2\pi} \dot{\phi}(k) dk,
\end{equation}
where
\begin{eqnarray}\label{phi-dot}
 \dot{\phi}(k)=\frac{\partial\phi(k)}{\partial k}=\frac{1}{(d(k))^2}\sum_{ij}\epsilon_{ij}d^i(k)\frac{\partial d^j(k)}{\partial k}.
\end{eqnarray}
The curve in the plane described by the point $\gamma(k)=(d^1(k), d^2(k))$ can wind around the origin multiple times as $k$ traverses the Brillouin zone (BZ), giving rise to the notion of the winding number as an integer, $\delta \in \mathbb{Z}$. In this way, the homotopy classification of Hamiltonians in Eq.(\ref{chiral-bloch-ham}) reduces to the homotopy classification of maps from the BZ to the space $\RR^2$ with the origin removed, since $d(k) \neq 0$ is required for the model to be insulating. The value of $\delta$ is stable under any deformation of the Hamiltonian that preserves chiral symmetry and does not close the gap~\cite{avron}. 
For the deformations considered in this work, these correspond to smooth variations of the hopping parameters $u_j$ entering Eq.~(\ref{lr-ssh-equations}), which explicitly preserve the symmetries of the model.

\begin{figure}[htb]
\begin{center}
\includegraphics[scale=.45]{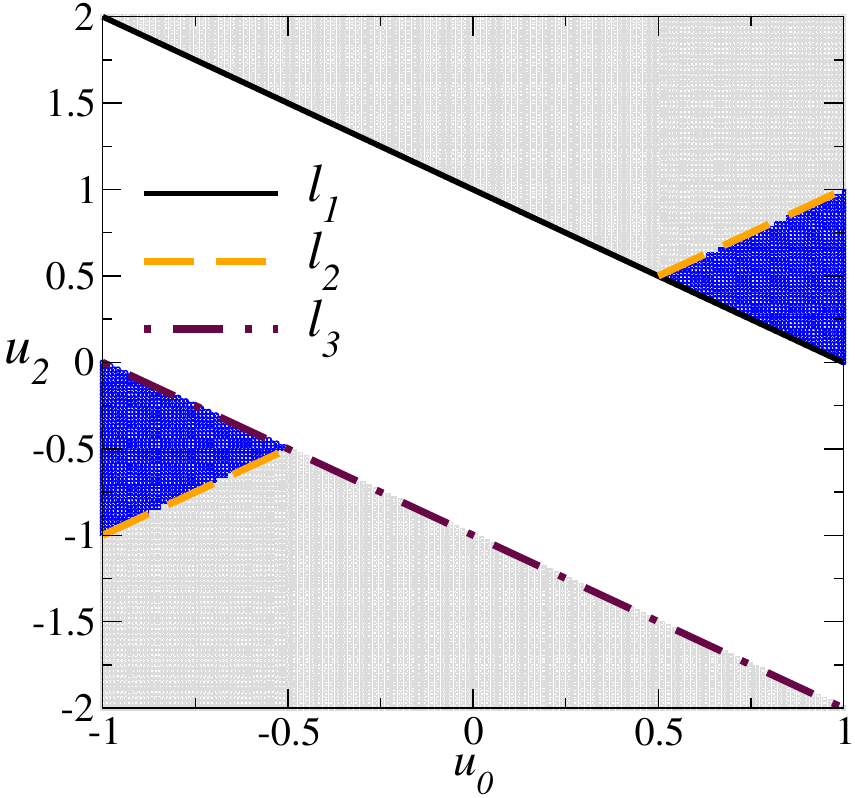}
\end{center}
\caption{(Color online) Phase diagram characterized by $\delta$ for the eSSH model of Eq. \ref{lr-ssh-equations}, as a function of the parameters $u_0$ and $u_2$, with $u_1=1$. The lines $l_1=1-u_0$, $l_2=u_0$, and $l_3=-1-u_0$ separate the phases with $\delta=0$ (blue region), $\delta=1$ (white region), and $\delta=2$ (grey region).}
\label{mapa-win}
\end{figure}

\begin{figure}[htb]
\begin{center}
\includegraphics[scale=.20]{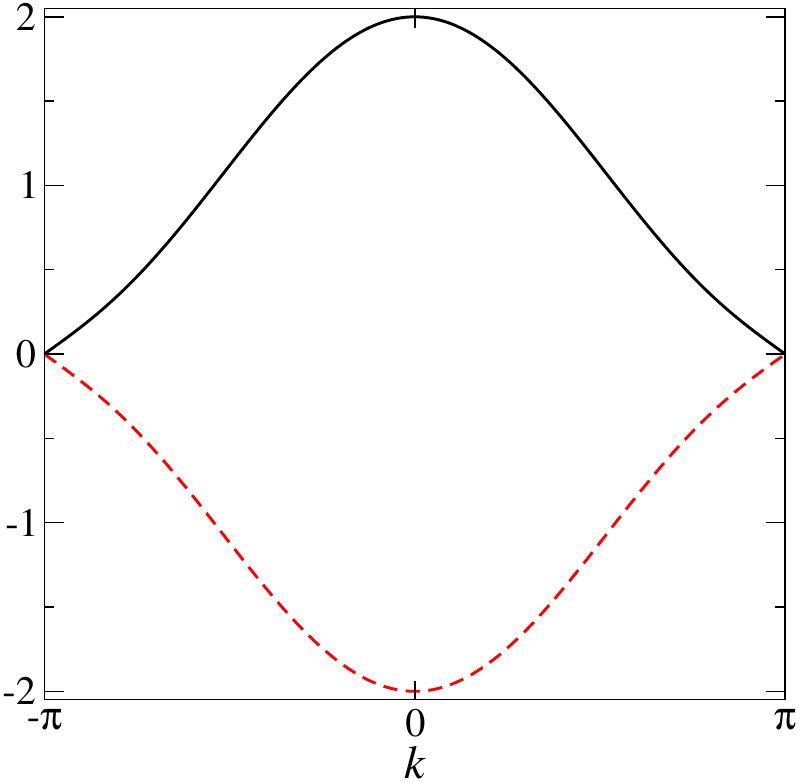}
\includegraphics[scale=.20]{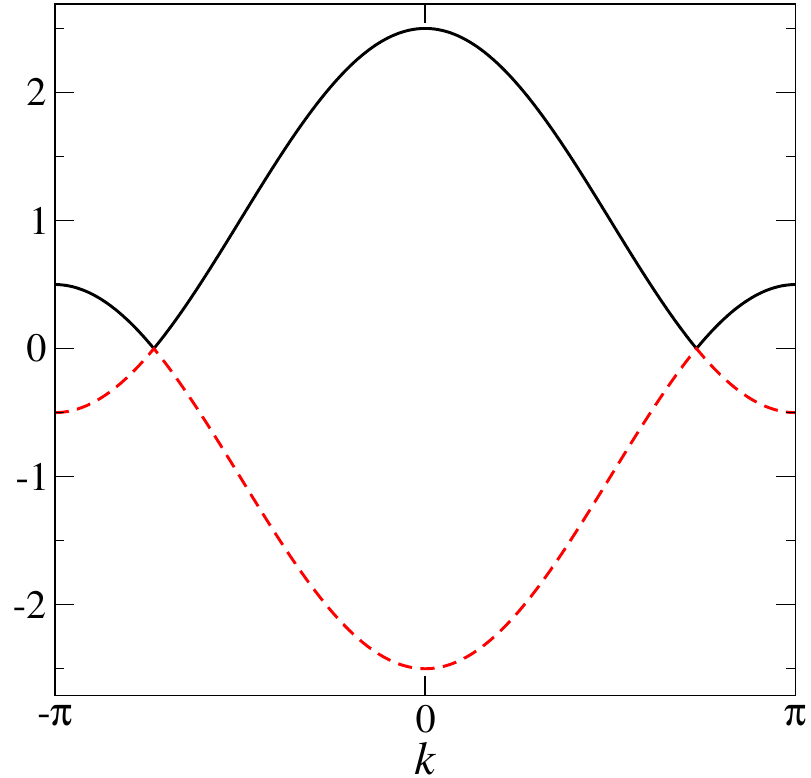}
\includegraphics[scale=.20]{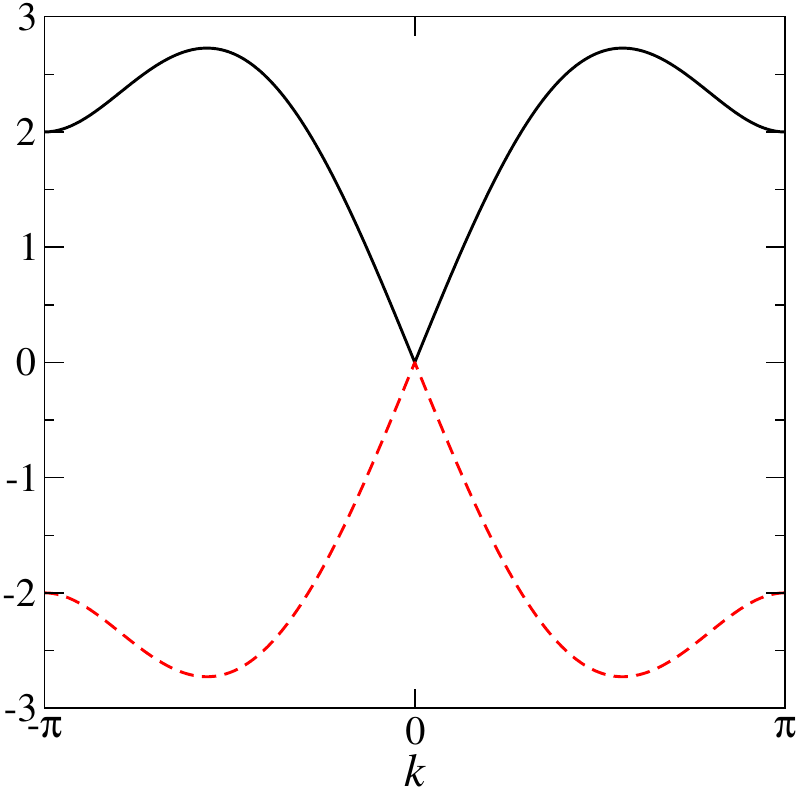}
\end{center}
\caption{(Color online) Band structure along the lines $l_1$ (left), $l_2$ (center) and $l_3$ (right) for $u_0/u_1 =0.75$.}
\label{band-structures}
\end{figure}

In what follows, we focus on the case $M=2$ ($u_j=0$ for $j>2$). In Fig.~\ref{mapa-win}, we show the phase diagram constructed from the values of \(\delta\) as a function of the model parameters, expressed
in units of \(u_1\). Along the boundary of each region, the lines \(l_i\), with \(i\in\{1,2,3\}\),  the energy gap closes at some points of the Brillouin zone, as shown in the example in Fig.~\ref{band-structures}.

\begin{figure}[htb]
	\begin{center}
		\includegraphics[scale=.45]{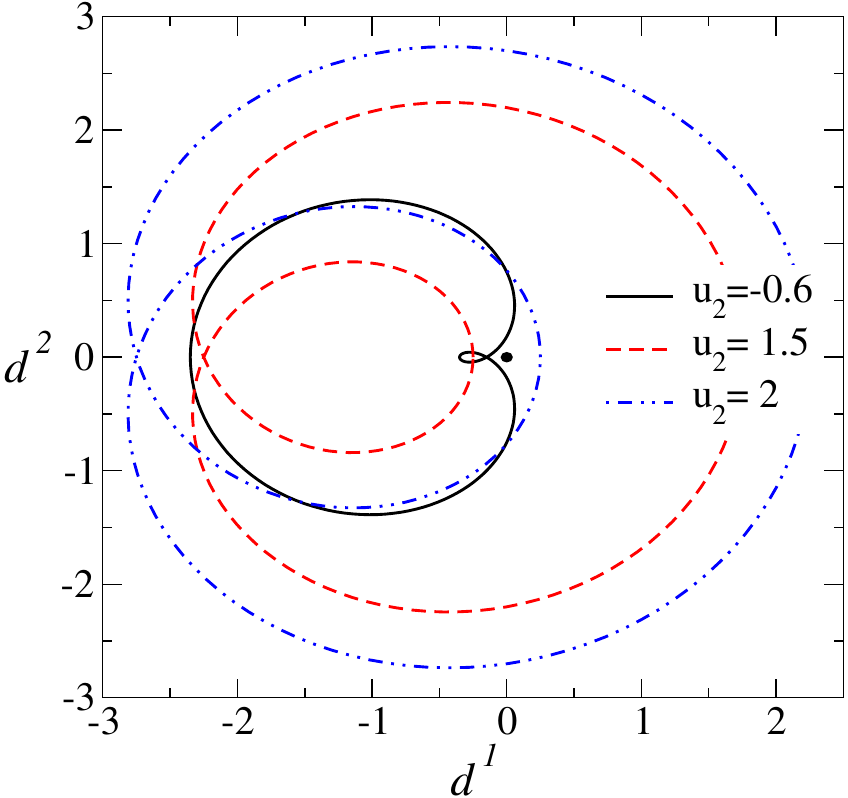}
		\includegraphics[scale=.45]{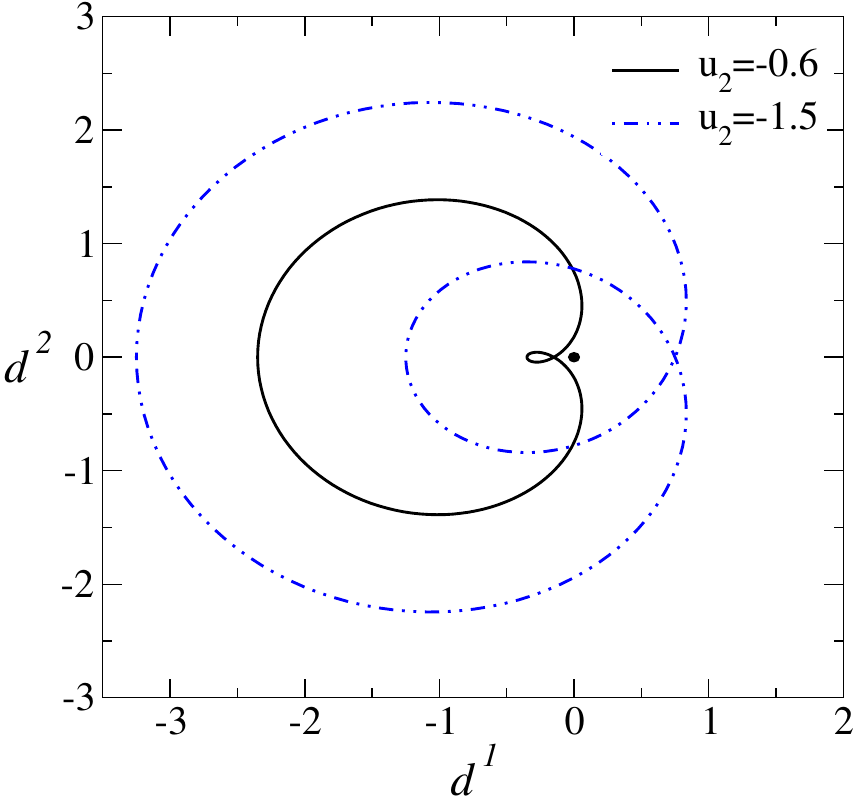}
	\end{center}
	\caption{(Color online) Top panel: $\gamma$ curves for $u_0=-0.75$, and $u_1=1$  and $u_2=2$ (dashed-dot-dot blue line), $u_2=1.5$ (dashed red line), and $u_2=-0.6$ (solid black line) with winding numbers $\delta=2$, $\delta=1$, and  $\delta=0$ respectively. Lower panel:  $\gamma$ curves for $u_2=-1.5$ (dashed-dot-dot blue line) and $u_2=-0.6$ (solid black line) with winding numbers $\delta=2$, and  $\delta=0$ respectively. Other parameters are the same that in the top panel. }
	\label{loops}
\end{figure}

Note that when the next-nearest-neighbour hopping $u_2$ dominates, the winding number is $\delta = 2$. In fact, due to the presence of the terms $\cos(2k)$ and $\sin(2k)$, for sufficiently large $|u_2|$ the image of the map $\gamma$ generally consists of two loops. Consequently, the winding number (which can take the values 0, 1, or 2) is governed by the relative configuration of these two loops, including whether they are concentric, mutually displaced, or of different radii.

The deformation processes starting from a curve with winding number $\delta = 2$, passing through a configuration with $\delta = 1$, and finally reaching $\delta = 0$ are exemplified by the path at constant $u_0 = -0.75$ and $2 > u_2 > -0.5$ in the phase diagram shown in Fig.~\ref{mapa-win}. The curves $\gamma$ for selected values of $u_2$ illustrating these transitions are shown in the top panel of Fig.~\ref{loops}.  

However, direct transitions from $\delta = 2$ to $\delta = 0$, without passing through a phase with $\delta = 1$, are also allowed. The path at constant $u_0 = -0.75$ and $-1.5 < u_2 < -0.5$ in the phase diagram provides an example. The corresponding curves $\gamma$ are shown in the lower panel of Fig.~\ref{loops} for two selected values of $u_2$. 

Specifically, from the analysis of the next Section, choosing
without loss of generality $u_0,u_1>0$, we find in agreement with
Fig.~\ref{mapa-win} two bounds states at each end of a finite chain for $u_2<-u_0-u_1$ (negative $u_2$). For positive $u_2$, two bound states appear in the following two cases: (i) $u_2>u_1-u_0$ for $u_1>2u_0$ and (ii) $u_2>u_0$ for $u_1<2u_0$.

\section{Semi-infinite chain and edge states}\label{section-edge-states}

In this section, we study the zero-energy states of the semi-infinite eSSH model with $M=2$. We follow the method of Alase \textit{et al.}  as presented in Refs. \cite{alase-1, alase-2, aligia-1}.
This method is particularly advantageous when the energy
of the eigenstates $E$ is known a priori, in particular for $E=0$ as is the case for topological end states \cite{aligia-1}. For the case $M=1$ this method has been used in Ref. \cite{aligia-2} and extended for finite long chains providing analytical results in very good agreement with numerical ones.

The Hamiltonian is given by:
\begin{eqnarray}\label{semi-infinite-chain}
 H&=&\sum_{j=1}^{\infty} u_0| B_j\rangle\langle A_j| + u_1| B_j\rangle\langle A_{j+1}| +
   u_2| B_j\rangle\langle A_{j+2}| \nonumber\\
  &+& \mbox{H.c.},
\end{eqnarray}
By a convenient change of sign of the states, we
assume $u_0>0$ and $u_1>0$ without loss of generality.

Associated with this Hamiltonian, we define the projector operators onto the bulk, $P_b$, and the edge,  $P_e$, as follows
\begin{equation}\label{projectors}
\begin{array}{l}
P_b = \sum_{j=3}^{\infty}P_j\\[8pt]
P_e = P_1+P_2,
\end{array}
\end{equation}
with $P_j=|A_j\rangle\langle A_j|+|B_j\rangle\langle B_j|$. By definition $P_b+P_e=1$.
The origin of this separation can be understood as follows. Deep in the bulk, all hopping processes (both to the left and to the right) are allowed, whereas this is no longer the case near the boundary. The projector onto the bulk $P_b$ excludes sites close to the boundary that would otherwise involve hoppings outside the system. This projection yields a locally translationally invariant Hamiltonian, allowing one, as we show below, to solve the projected
Schr\"{o}dinger equation in terms of states analogous to Bloch waves, but with a complex wave vector
[see Eqs. (\ref{z-state})].
The main idea is to split the Schr\"{o}dinger equation,
$H|\psi\rangle = E|\psi\rangle$, for a given state $|\psi\rangle$ of energy $E$, in two parts, one for the bulk and the other for the edge \cite{note2}.
\begin{equation}\label{bulk-edge-problem-0}
\begin{array}{l}
P_b H |\psi\rangle = EP_b  |\psi\rangle\\[8pt]
P_e H |\psi\rangle = EP_e  |\psi\rangle.
\end{array}
\end{equation}
In particular,
\begin{eqnarray}\label{bulk-edge-problem-1}
P_e H &=& u_0\big( | A_1\rangle\langle B_1| + | B_1\rangle\langle A_1| + | A_2\rangle\langle B_2| + | B_2\rangle\langle A_2|\big) + \nonumber\\ && u_1\big(| B_1\rangle\langle A_2|+| A_2\rangle\langle B_1|+| B_2\rangle\langle A_3|\big)+\nonumber\\ && u_2\big(| B_1\rangle\langle A_3|+| B_2\rangle\langle A_4|\big)\nonumber\\
P_b H &=& H-P_e H.
\end{eqnarray}

We aim to solve the bulk eigenvalue equation of the Eq. (\ref{bulk-edge-problem-0}). To this end, we use generalized Bloch states expanded in powers of a complex number $z$,
\begin{equation}\label{z-state}
\begin{array}{l}
|\psi_{A}(z)\rangle = \sum_{j=1}^{\infty} z^{j-1}|A_j\rangle\\[8pt]
|\psi_{B}(z)\rangle = \sum_{j=1}^{\infty} z^{j-1}|B_j\rangle.
\end{array}
\end{equation}
Note that
\begin{eqnarray}\label{bulk-edge-problem-2}
H |\psi_{A}(z)\rangle =&& \sum_{j=1}^{\infty} z^{j-1} H|A_j\rangle\nonumber\\
   =&& u_0\sum_{j=1}^{\infty} z^{j-1}| B_j\rangle + u_1\sum_{j=2}^{\infty} z^{j-1}| B_{j-1}\rangle+\nonumber\\ &&
   u_2\sum_{j=3}^{\infty} z^{j-1}| B_{j-2}\rangle\nonumber\\
   =&& \big( u_0  + zu_1  + z^2 u_2 \big)|\psi_{B}(z)\rangle.
\end{eqnarray}
However,
\begin{eqnarray}\label{bulk-edge-problem-3}
H |\psi_{B}(z)\rangle =&& \sum_{j=1}^{\infty} z^{j-1} H|B_j\rangle\nonumber\\
=&& u_0\sum_{j=1}^{\infty} z^{j-1}| A_j\rangle + u_1\sum_{j=1}^{\infty} z^{j-1}| A_{j+1}\rangle +\nonumber\\
   &&u_2\sum_{j=1}^{\infty} z^{j-1}| A_{j+2}\rangle\nonumber\\
   =&&\big( u_0  +\frac{u_1}{z}  + \frac{u_2}{z^2} \big)|\psi_{A}(z)\rangle-\nonumber\\&&\big( \frac{u_1}{z}  + \frac{u_2}{z^2} \big)|A_1\rangle-\frac{u_2}{z}|A_2\rangle.
\end{eqnarray}
By applying $P_b$ to the previous equation we obtain:
\begin{eqnarray}\label{bulk-edge-problem-3b}
P_b H |\psi_{B}(z)\rangle=&& \big( u_0+\frac{u_1}{z}+\frac{u_2}{z^2} \big)P_b|\psi_{A}(z)\rangle-\nonumber\\&&\big( \frac{u_1}{z}  + \frac{u_2}{z^2} \big)P_b|A_1\rangle-\frac{u_2}{z}P_b|A_2\rangle\nonumber\\
=&&\big( u_0+\frac{u_1}{z}+\frac{u_2}{z^2} \big)P_b|\psi_{A}(z)\rangle.
\end{eqnarray}
The idea behind applying $P_b$ is to obtain equations                                                                                                                                                                                                             similar to those corresponding to a Bloch Hamiltonian $H(k)$ (with $z$ playing the role of $e^{-ik}$) despite the absence of periodic boundary conditions.
Note that, due to the presence of $u_2$ in the Hamiltonian, the edge must include the cell $j=2$ in order to prevent $P_bH$ from acquiring a correction arising from a fictitious hopping between this cell and the non-existent cell $j=0$.
We therefore arrive at the following set of two coupled equations
\begin{equation}\label{bulk-edge-problem-3c}
\begin{array}{l}
P_b H |\psi_{A}(z)\rangle= \big( u_0  + zu_1  + z^2 u_2 \big)P_b|\psi_{B}(z)\rangle\\[8pt]
P_b H |\psi_{B}(z)\rangle= \big( u_0+\frac{u_1}{z}+\frac{u_2}{z^2} \big)P_b|\psi_{A}(z)\rangle.
\end{array}
\end{equation}
Let $|\psi\rangle$ be $|\psi\rangle=u(z)|\psi_{A}(z)\rangle+v(z)|\psi_{B}(z)\rangle$. The bulk eigenvalue equation Eq. (\ref{bulk-edge-problem-0}), leads to the following matrix equation for the coefficients $u(z)$ and $v(z)$:
\begin{eqnarray}\label{bulk-edge-problem-4}
 \left(\begin{array}{cc}
        -E & Q_2(z)\\
       Q_1(z) & -E
      \end{array}\right)\left(\begin{array}{c}
        u(z) \\
        v(z) \end{array}\right)=\left(\begin{array}{c}
        0 \\
        0 \end{array}\right),
\end{eqnarray}
with $Q_1(z)=u_0  + zu_1  + z^2 u_2$ and $Q_2(z)=u_0  +\frac{u_1}{z}  + \frac{u_2}{z^2}$.
There is non-trivial solution if and only if
\begin{eqnarray}\label{bulk-edge-problem-5}
 E^2 -Q_1(z)Q_2(z)=0.
\end{eqnarray}

In general, $E$ should be determined selfconsistently using the equation for the edge.
However, the procedure is greatly simplified if one looks for zero-energy eigenstates \cite{aligia-1}.
For $E=0$ we have
\begin{eqnarray}\label{bulk-edge-problem-6}
 Q_1(z)Q_2(z)=0.
\end{eqnarray}

Furthermore, the edge contribution of Eq. (\ref{bulk-edge-problem-0}), $P_e H |\psi\rangle=0$, becomes
\begin{eqnarray}\label{bulk-edge-problem-6b}
0=&&u(z)Q_1(z)\Big(|B_1\rangle+z|B_2\rangle\Big)+\nonumber\\&& v(z)\Big(u_0|A_1\rangle+\big( z Q_2(z) - \frac{u_2}{z} \big)|A_2\rangle\Big).
\end{eqnarray}

Equation (\ref{bulk-edge-problem-6}) is satisfied if one of the two factors, $Q_1(z)$ or $Q_2(z)$, vanishes. The procedure to obtain the correct roots is as follows: one solves one the factors and obtains two roots, $z_{\pm}$. If $ z_i  \ne 0$,  for $i=\pm$, then $z^{-1}_i$ is a root of the other factor. To obtain states localized at the left end of the chain it is required that
$\arrowvert z_i \arrowvert <1$ \cite{note}

The two roots of the first factor, $Q_1(z)$, are:
\begin{eqnarray}\label{bulk-edge-problem-7}
 z_{\pm}=\frac{-1\pm\sqrt{1-4\tilde{u}_0\tilde{u}_2}}{2\tilde{u}_2},
\end{eqnarray}
where $\tilde{u}_0=u_0/u_1$ and $\tilde{u}_2=u_2/u_1$.

Note that upon replacing $z$ by $e^{-ik}$, the function $Q_1(z)$
has the same for as $Q(k)$ entering Eq. (\ref{chiral-bloch-ham}). Therefore if $k$ is a solution of $Q(k)=0$,
indicating a closing of the gap of the Bloch Hamiltonian, then $z=e^{-ik}$ is a solution of $Q_1(z)=0$ with $|z|=1$. As the parameters are varied, a change in the winding number occurs when $Q(k)=0$, while a change in the number
of localized end states takes place when $|z|=1$. These arguments show that the two phenomena are directly related, in agreement with the bulk–boundary correspondence. This reasoning remains valid for $M>2$.

In what follows, we analyze different cases:

(i). $\tilde{u}_2$ is large enough and positive.
In this case, both roots are complex conjugate: $z_{\pm}=-\frac{1}{2\tilde{u}_2}\pm i\sqrt{\frac{\tilde{u}_0}{\tilde{u}_2}-(\frac{1}{2\tilde{u}_2})^2}$, with $\arrowvert z_+ \arrowvert = \arrowvert z_- \arrowvert =\sqrt{u_0/u_2}<1$. From Eq. (\ref{bulk-edge-problem-4}), we find that $v(z_\pm)=0$, and there are two solutions of the form $|\psi_{A}(z_\pm)\rangle$. Note that the edge equation [second Eq. (\ref{bulk-edge-problem-0})] is satisfied. Therefore, we obtain a pair of localized states of the form of the first Eq. (\ref{z-state}), whose weight lies on the $A$-type sites only.

An important case is the one in which the root \(z\) tends to a real value.
This occurs along the line \(\tilde{u}_2 = \frac{1}{4\tilde{u}_0}\), which is plotted in Fig.~\ref{wind-li2} as a red dashed line and becomes relevant for the analysis of the experiment discussed in Sec.~\ref{subsec-Sli}.
At first sight, it appears that there is only one real solution, corresponding to a single edge state.
However, as can be seen from Fig.~\ref{wind-li2}, the line \(\tilde{u}_2 = \frac{1}{4\tilde{u}_0}\) is consistent with either zero or two edge states.
Writing \(z_i = |z| \exp(\pm i \theta)\), the limit \(\theta \to 0^{+}\) with \(|\tilde{u}_2|>1/2\) and \(|\tilde{u}_0|<1/2\), which leads to \(|z|<1\), allows for two orthonormalized edge states. If, however, \(|\tilde{u}_2|<1/2\) and \(|\tilde{u}_0|>1/2\) where \(|z|>1\), there are no edge states.
This limit is taken explicitly in Appendix~\ref{appendix-theta0}.

(ii). $\tilde{u}_2$ is large enough and negative. the roots are $z_{\pm}=\frac{1}{2|\tilde{u}_2|}\mp \sqrt{\frac{\tilde{u}_0}{|\tilde{u}_2|}+(\frac{1}{2|\tilde{u}_2|})^2}$, again with $\arrowvert z_{\pm} \arrowvert<1$ with two solutions of the form $|\psi_{A}(z_\pm)\rangle$. Again, Eq. (\ref{bulk-edge-problem-0}) is satisfied.

Note that, according to the bulk-boundary correspondence, the appearance of two edge states in the cases (i) and (ii), is consistent with the bulk winding number
$\delta=2$, see the upper grey region of figure \ref{mapa-win}.

(iii). $\tilde{u}_2$ is small. From Eq. (\ref{bulk-edge-problem-7}) we obtain:
\begin{eqnarray}\label{bulk-edge-problem-8}
 z_{\pm}&=&\frac{-1\pm \sqrt{1-4\tilde{u}_0\tilde{u}_2}}{2\tilde{u}_2}\approx\frac{-1\pm (1-2\tilde{u}_0\tilde{u}_2)}{2\tilde{u}_2}
\end{eqnarray}
with solutions $z_+\approx-\tilde{u}_0$ and $z_-\approx-\frac{1}{\tilde{u}_2}$ in leading order. Note that within this approximation, $|z_-|>1$. If $\tilde{u}_0<1$, there exists a solution $z_+$ that is closely analogous to that of the standard SSH model \cite{aligia-2}
for which $v(z_+)=0$. Thus, this solution corresponds to an edge state of the form
$|\psi_{A}(z_+)\rangle$.

In addition, $1/z_-$ is a root of $Q_2(z)$ with $|1/z_-|<1$, implying $u=0$. If this were an edge state, it
should be of the form $|\psi_{B}(z_-)\rangle$. However, the edge contribution of Eq. (\ref{bulk-edge-problem-6b}) is not satisfied.
Therefore, this regime is consistent with the bulk winding number $\delta=1$, see the white region of figure \ref{mapa-win}. If instead $\tilde{u}_0>1$, $|z_+| >1$, so there is no edge state.

In summary, due to the equation for the boundary, only solutions of the form
$|\psi_{A}(z_+)\rangle$ with $|z|<1$ satisfying $Q_1(z)=0$ are localized states of zero energy.

Special case: Within (ii) we have $|z_+|<|z_-|$. Thus, as $\tilde{u}_0$ increases, when $|z_-|=1$, one edge state disappears. This is consistent with the phase transition that occurs when crossing the $l_3$ line (grey to white) in the lower sector of Figure  \ref{mapa-win}. In particular, from $|z_-|=1$ we obtain $|\tilde{u}_2|=1+\tilde{u}_0$. Therefore, if $|{u}_2|>{u}_0+{u}_1$ there are two localized zero-energy states, and only one for $|{u}_2|<{u}_0+{u}_1$.

It is worth mentioning that, in more general cases including hopping $u_N$ up to the $N^{th}$ nearest neighbors of different atoms in the unit cell
($A \leftrightarrow B$), the function $Q_1$ defined in
Eq.~(\ref{bulk-edge-problem-4}) becomes a complex polynomial of degree $N$ with $N$ roots, allowing for the existence of up to $N$ edge states.

\section{Application to particular cases}\label{section-results}

Here, we analyze the structure and number of edge states for the model parameters representing the two recent experiments mentioned in the Introduction.

We will denote each set of parameters modeling each Hamiltonians $H(k)$ by $(u_0, u_1, u_2)$.
\subsection{Edge states in one-dimensional finite chiral photonic lattices}\label{subsec-Cli}

While topological phases and zero-energy edge states are strictly well defined only in the thermodynamic limit, experimental realizations and numerical calculations are often performed on finite systems. Nevertheless, for sufficiently large systems, well-localized boundary states and a small energy gap emerge, from which the behavior in the thermodynamic limit can be reliably inferred.

In Reference~\cite{li}, the authors obtain the number of topological edge-states by diagonalizing the matrix Hamiltonian describing a $32$-sites eSSH chain.
In particular, for the case of a model including second-neighbor hopping $u_{2}$, as shown in Fig.~\ref{finite-chain-cartoon}, the phase diagram is shown in Fig. \ref{mapa-win}.

The authors analyze the following sets of parameters:
$\check{1} = (1, 1.5, 4.8)$ and $\check{2} = (1, 0.6, 4.8)$,
which correspond to points A and B in Fig.~2(b) of Ref. \cite{li} respectively.
The results for the intensity as a function of site are represented in
Figs.~3(a) and 3(b) of Ref. \cite{li} respectively.
In both cases, $u_2 > u_0$ and $u_2 > 1 - u_0$, for which, according to the phase diagram the winding number is $\delta = 2$.
Briefly, their results exhibit, in both cases, two pairs of low-energy states.
Considering only those of positive energy, we label them as I and II. State~I is characterized by having its maximum weight on the first site, $A_1$ in figure \ref{finite-chain-cartoon}, with a smaller contribution located inside the chain. In contrast, state~II has its maximum weight on the third site, $A_2$ in figure \ref{finite-chain-cartoon}.

As expected, the zero-energy states built in Section~\ref{section-edge-states} for semi-infinite chains explain the features of the aforementioned states~I and~II.
By applying the results derived in Eq.~\ref{bulk-edge-problem-7}, we obtain
$z_{\check{1},\pm} = \frac{\sqrt{30}}{12}e^{\pm i\theta_{\check{1}}}$ and
$z_{\check{2},\pm} = \frac{\sqrt{30}}{12}e^{\pm i\theta_{\check{1}}}$
for the parameter sets $\check{1}$ and $\check{2}$, respectively, with
$\theta_{\check{1}}=\pi-\arctan(\sqrt{113/15}) \approx 1.92$ and $\theta_{\check{2}}=\pi-\arctan(\sqrt{157/3}) \approx 1.71$. Therefore, in both cases we obtain two zero-energy edge states of the form $|\psi_{A}(z_{\pm})\rangle$.
Since, for a finite chain described by a real Hamiltonian, the eigenstates are real, it is more convenient to change the basis as explained in Appendix~\ref{appendix-theta0}. We therefore introduce the two orthonormal real states given in Eqs.~(\ref{d1}) and (\ref{s-prime2}). In this basis, $|\widehat{\psi_{-}}(\check{1})\rangle$ is proportional to the difference of the states $|\psi_{A}(z_{\pm})\rangle$,
and remains well defined in the limit in which $z_{+}$ and $z_{-}$ approach to
the same real value,

\begin{eqnarray}\label{estados-check-1}
|\widehat{\psi_{-}}(\check{1})\rangle &=& 1.005\sum_{j=1}^{\infty}|z|^{j-1}\sin(j\phi)|A_{j+1}\rangle\nonumber\\
|\widehat{\psi'}(\check{1})\rangle&=&1.064\sum_{j=1}^{\infty}|z|^{j-1}\cos((j-1)\phi)|A_{j}\rangle\nonumber\\
&&-0.66\sum_{j=1}^{\infty}|z|^{j-1}\sin(j\phi)|A_{j+1}\rangle,
\end{eqnarray}
and
\begin{eqnarray}\label{estados-check-2}
|\widehat{\psi_{-}}(\check{2})\rangle &=& 0.985\sum_{j=1}^{\infty}|z|^{j-1}\sin(j\phi)|A_{j+1}\rangle\nonumber\\
|\widehat{\psi'}(\check{2})\rangle&=&0.955\sum_{j=1}^{\infty}|z|^{j-1}\cos((j-1)\phi)|A_{j}\rangle\nonumber\\
&&-0.51\sum_{j=1}^{\infty}|z|^{j-1}\sin(j\phi)|A_{j+1}\rangle,
\end{eqnarray}
for the sets $\check{1}$ and $\check{2}$ respectively.
For the numerical solution of a finite chain, one expects that the resulting low-energy states are similar to the above real orthogonal states,
except for the fact that the numerical states contain a mixture of end
states decaying to the left and to the right (as we show below),
while by construction
the analytical states correspond to a semi-infinite chain and are localized only at one end.

\begin{figure}[htb]
\begin{center}
\includegraphics[scale=.45]{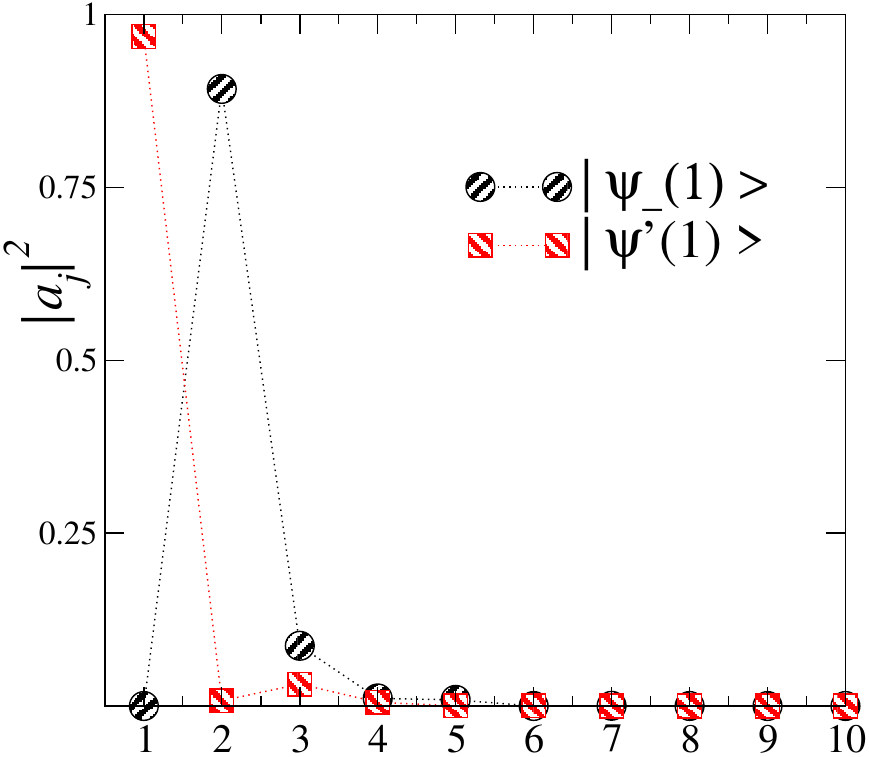}
\includegraphics[scale=.45]{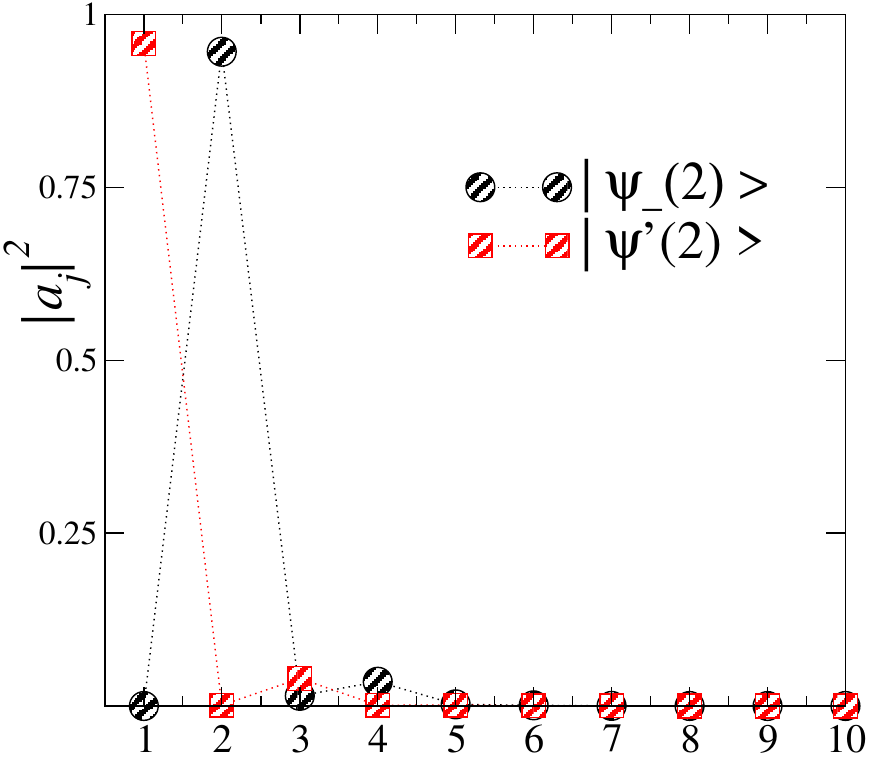}
\end{center}
\caption{(Color online) Weights, $\vert a_j \vert ^2$, of the edge states for the sets $\hat{1}$ and $\hat{2}$ calculated from Eq.~(\ref{z-state}). The horizontal axis runs over cells.}
\label{cli}
\end{figure}

The weights of each state,
$\vert a_j \vert ^2$, in each set of parameters, are shown in Fig.~\ref{cli}. Note that in both cases, set $\check{1}$ and set $\check{2}$, the state $|\widehat{\psi'}\rangle$ has its maximum weight over the site $A_1$(first site) and exhibits a small contribution coming from $A_4$ ($7^{th}$ site), clearly visible in the set
$\check{2}$. In contrast, the state $|\widehat{\psi_{-}}\rangle$ has no contribution coming from $A_1$. It has its maximum weight over the site $A_2$ (third site).

It follows that the numerically obtained states I and II in Ref.~\cite{li} have the structure of the states $|\widehat{\psi'}\rangle$ and $|\widehat{\psi_{-}}\rangle$ respectively.

\begin{figure}[htb]
	\begin{center}
		\includegraphics[scale=.45]{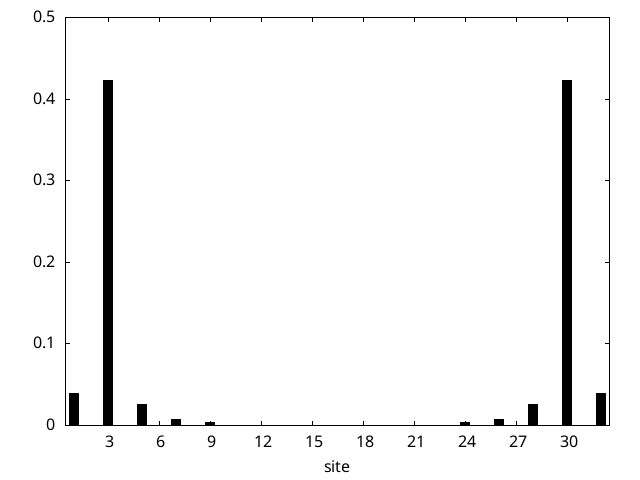}
		\includegraphics[scale=.45]{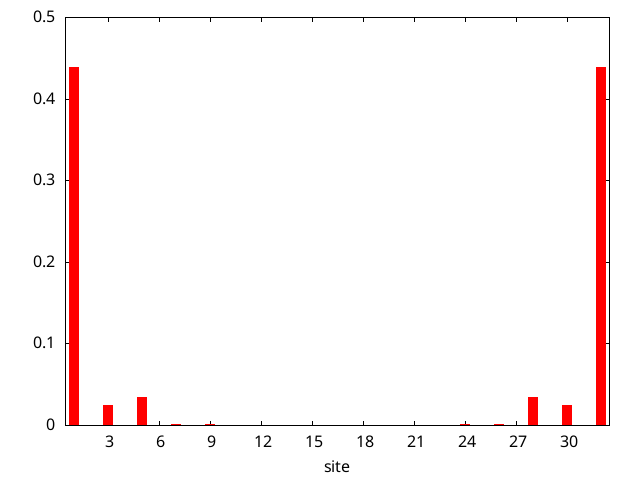}
	\end{center}
	\caption{(Color online) Weights, $\vert a_j \vert ^2$, of the edge states I (top panel) and II (bottom panel) as a function of lattice site for the set $\hat{1}$ with $N=16$.}
	\label{cli-1-n16}
\end{figure}

\begin{figure}[htb]
	\begin{center}
		\includegraphics[scale=.45]{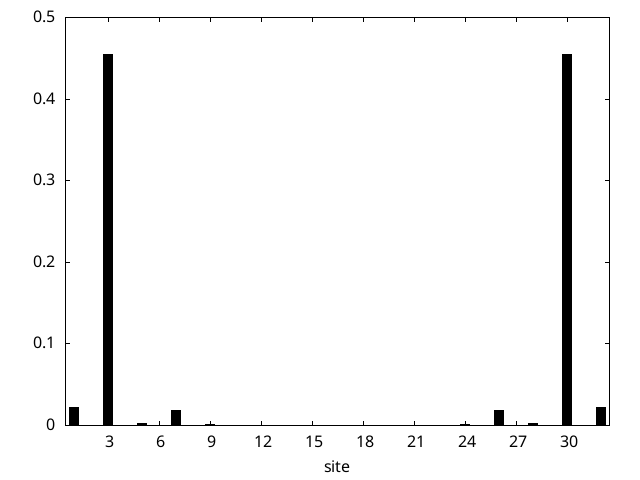}
		\includegraphics[scale=.45]{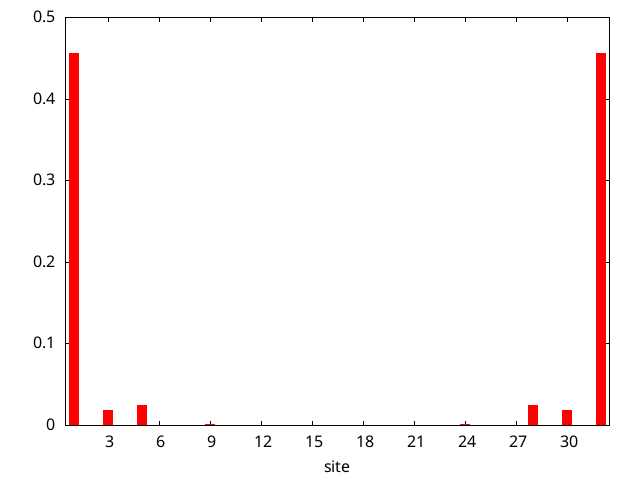}
	\end{center}
	\caption{(Color online) Same as Fig. \ref{cli-1-n16} for the set $\hat{2}$.}
	\label{cli-2-n16}
\end{figure}

Moreover, the exact analytical results obtained for a semi-infinite chain in Sec.~\ref{section-edge-states} can be extended to the case of a finite chain
of $N$ unit cells within an error of the order of $|z|^{2N}/\Delta_g$ in the energy, where
$\Delta_g$ is the gap of the chain with PBC \cite{aligia-2}.
The steps required to do so are outlined in Appendix~\ref{appendix-finite-chain}. Using these results for $N=16$, we obtain
two pairs of edge states, each with its weight equally distributed between the two ends of the chain. The weight distribution of each pair is shown in Figures \ref{cli-1-n16} and \ref{cli-2-n16} for the sets of parameters
$\hat{1}$ and $\hat{2}$ respectively.
We obtain the energies $E_I=2.56 \times 10^{-5}$ and  $E_{II}=9.69 \times 10^{-6}$ for the states I and II for the parameters of the set $\hat{1}$ and
$E_I=3.07 \times 10^{-5}$ and  $E_{II}=8.55 \times 10^{-6}$ for the
set $\hat{2}$. In both cases $|z|=0.4564$ and $|z|^{16}=3.54 \times 10^{-6}$. Furthermore, the value of $\Delta_g$ [the minimum value of $|Q(k)|$ in Eq. (\ref{chiral-bloch-ham})]
is $7.14$ and $2.53$ for the sets  $\hat{1}$ and $\hat{2}$ respectively.

In each case, the states are superpositions of the finite-chain versions of
$|\widehat{\psi_{-}}\rangle$ and $|\widehat{\psi'}\rangle$. It is clear
that the member of each pair displayed in the top panels of Figs.~\ref{cli-1-n16} and \ref{cli-2-n16} has a dominant contribution from
$|\widehat{\psi_{-}}\rangle$, exhibiting its maximum weight at the third site, $A_2$. Similarly, the other member of the pair, shown in the lower panels of
Figs.~\ref{cli-1-n16} and \ref{cli-2-n16}, closely follows the spatial profile of
$|\widehat{\psi'}\rangle$.

Note that the state of highest energy is related
to $|\widehat{\psi_{-}}\rangle$, the difference between $|\psi_{A}(z_+)\rangle$
and $|\psi_{A}(z_-)\rangle$.
This behavior can be understood as follows. When restricted to the low-energy sector, these two states (and the B-type ones) span a $4 \times 4$ matrix of $H$ with a non-zero overlap $S$. From the well-known solution of this problem in the large-overlap regime
(in particular for small $|z|$ or when $z_+$ and $z_-$ are very similar),
specifically for $S \rightarrow 1$, the higher-energy solution corresponds to the antisymmetric combination of the two states.

In any case, there is an excellent agreement with the numerical results presented in Reference~\cite{li}, but with the advantage of a well understanding of the weight distribution of each edge state.
\subsection{Edge states in one-dimensional bipartite superradiance lattices}\label{subsec-Sli}

Topological phase transitions in atomic ensembles generating one-dimensional bipartite superradiance lattices were studied both experimentally and theoretically by Shuai Li and collaborators in Reference \cite{li2}. In particular, the authors report a topological phase transition, tunable by a parameter $\eta$ defined below, between phases characterized by winding numbers $\delta = 0$ and $\delta = 2$, with $\delta$ calculated from the effective bulk Hamiltonian [see Eq.~(\ref{li2-bloch-ham}) below and Eq.~(\ref{delta})]. From the experimental point of view, the transition is inferred from changes in the peaks of the superradiance spectra. Here, we apply the machinery developed in Section~\ref{section-edge-states} to obtain the edge states along the topological phase transition.
The Hamiltonian model describing the superradiance lattice is given by (see Eq. (4) of Ref. \cite{li2}),
\begin{eqnarray}\label{li2-bloch-ham}
 \tilde{H_s}(k)=\left(\begin{array}{cc}
        d^0(k) & d^1(k)-id^2(k)\\
        d^1(k)+id^2(k) & d^0(k)
      \end{array}\right),
\end{eqnarray}
where the real functions $d^0(k)$, and $d^1(k)$, and $d^2(k)$ are given by:
\begin{equation}\label{li2-equations}
\begin{array}{l}
d^0(k) = \Omega^2+\tilde{\Omega}^2+2\Omega\tilde{\Omega}\mbox{cos}(k) \\[8pt]
d^1(k) = \tilde{\Omega}^2+2\Omega\tilde{\Omega}\mbox{cos}(k)+\Omega^2\mbox{cos}(2k)\\[8pt]
d^2(k) = 2\Omega\tilde{\Omega}\mbox{sin}(k)+\Omega^2\mbox{sin}(2k).
\end{array}
\end{equation}
where $\Omega$ and $\tilde{\Omega}$ are Rabi frequencies in units of some characteristic detuning frequency $\Delta$. Since the diagonal component does not affect the topological properties of the model we will not consider it. Note that $\tilde{H_s}(k)=d^0(k)I+H(k)$,  where $I$ is the $2\times 2$ identity matrix and the Hamiltonian $H(k)$ takes the form given by Eq. (\ref{chiral-bloch-ham}) with:
\begin{equation}\label{li2-components}
\begin{array}{l}
u_0(\eta) = 1\\[8pt]
u_1(\eta) = 2\eta\\[8pt]
u_2(\eta) = \eta^2.
\end{array}
\end{equation}
where $\eta=\Omega/\tilde{\Omega}\ne0$ is the only tunable parameter of the model. Without loss of generality, we have fixed $\tilde{\Omega}^2/\Delta=1$.
\begin{figure}[htb]\begin{center}
\includegraphics[scale=.30]{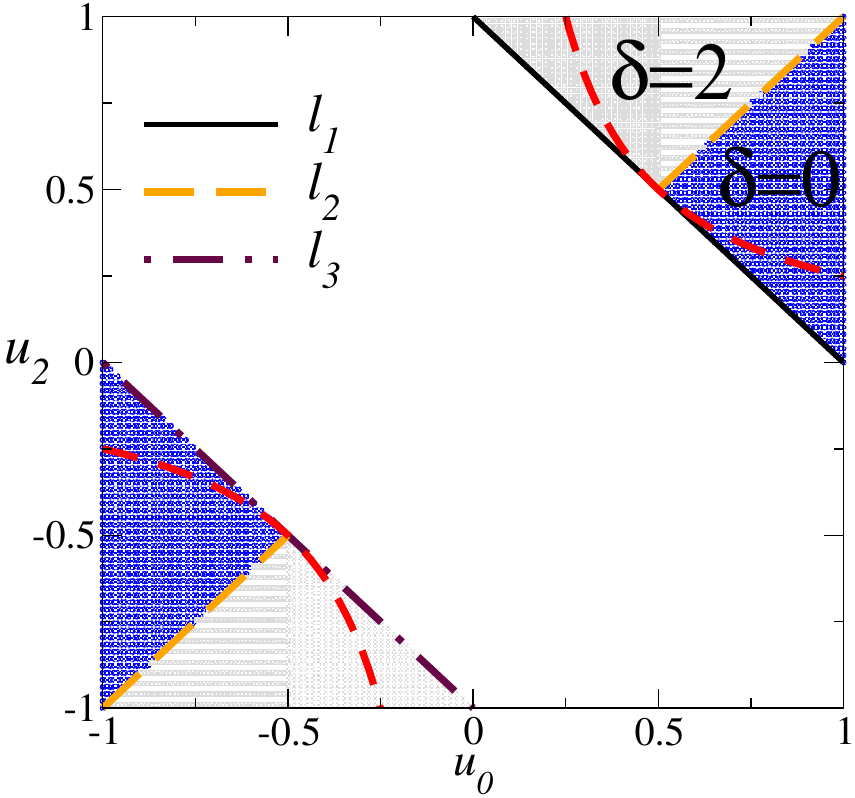}
\includegraphics[scale=.30]{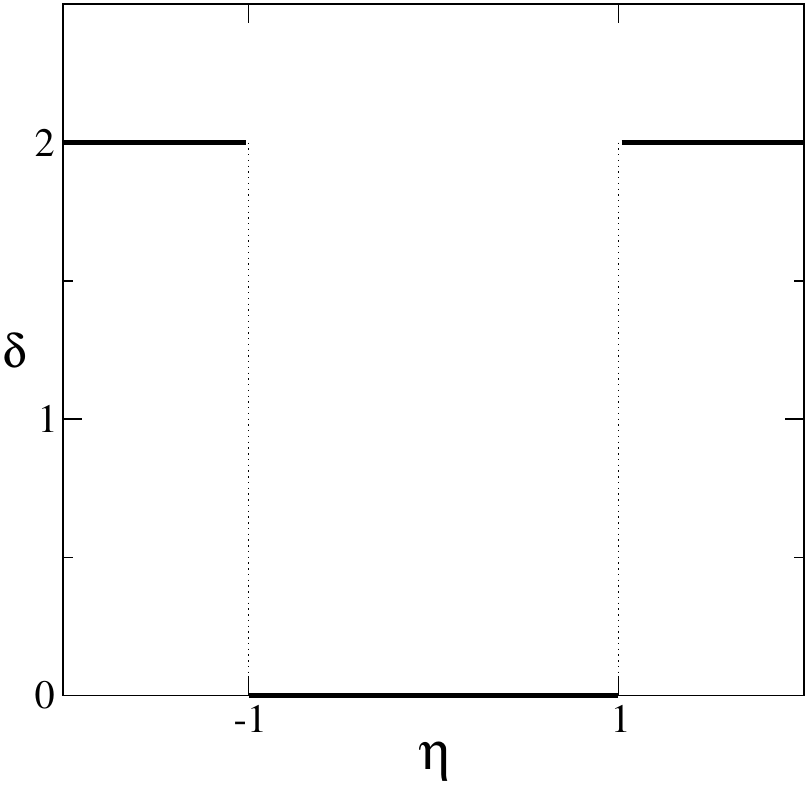}
\end{center}
\caption{(Color online) Left: Phase diagram of the superradiance lattice model for $u_1=1$. Colors have the same meaning as in Fig. \ref{mapa-win}. Right:
winding number as a function of $\eta$  along the line defined
by Eq. (\ref{li2-components}).}
\label{wind-li2}
\end{figure}

Note that, from Eq.~(\ref{li2-components}), the adiabatic changes of the Hamiltonian in Eq.~(\ref{li2-bloch-ham}) take place along the curve
$1 - 4 \tilde{u}_0(\eta)\tilde{u}_2(\eta) = 0$,
which is displayed in the left panel of Fig.~\ref{wind-li2} as a red dashed line. A topological phase transition from a trivial phase with $\delta = 0$ to a non-trivial phase with $\delta = 2$ is therefore expected. In fact, in the right panel of Fig.~\ref{wind-li2} we reproduce the results shown in Fig.~2(b) of Ref.~\cite{li2} for the winding number as a function of $\eta$. The aforementioned topological phase transition is clearly evident.

From Eq.~(\ref{bulk-edge-problem-7}) and Eq.~(\ref{li2-components}), we obtain $z_{+}=z_{-}=-\frac{1}{2\tilde{u}_2(\eta)}=-\frac{u_1(\eta)}{2u_2(\eta)}=-\frac{1}{\eta}$. So, when $|\eta|>1$ a pair of edge states are expected to be present at each border in a finite chain. In the case of a semi-infinite chain, in the Appendix \ref{appendix-theta0} [see Eq.~(\ref{d3}) and Eq.~(\ref{s3})] we show that such pair of states can be chosen to be:
\begin{eqnarray}\label{edge-states-li2-1}
|\widehat{\psi_{-}}(\eta)\rangle&=& \sum_{j=1}^{\infty} a^{-}_{j}(\eta)|A_{j+1}\rangle\nonumber\\
|\widehat{\psi_{+}}(\eta)\rangle&=& \sum_{j=1}^{\infty} a^{+}_{j}(\eta)|A_{j}\rangle,
\end{eqnarray}
where
\begin{eqnarray}\label{edge-states-li2-2}
a^{-}_{j}(\eta)&=& \sqrt{\frac{(\eta^2-1)^3}{\eta^2+1}}
 \frac{j}{|\eta|^{j+1}} \nonumber\\
a^{+}_{j}(\eta)&=&
 \frac{\sqrt{\eta^2-1}}{|\eta|^{j}}.
\end{eqnarray}
Note that $|\widehat{\psi_{+}}\rangle$, similar to the results presented in Sec.~\ref{subsec-Cli}, has maximum weight over $A_1$ while $|\widehat{\psi_{-}}\rangle$ over the following cells.

Contrary, when $|\eta|<1$, no edge state are present.

\begin{figure}[htb]\begin{center}
\includegraphics[scale=.28]{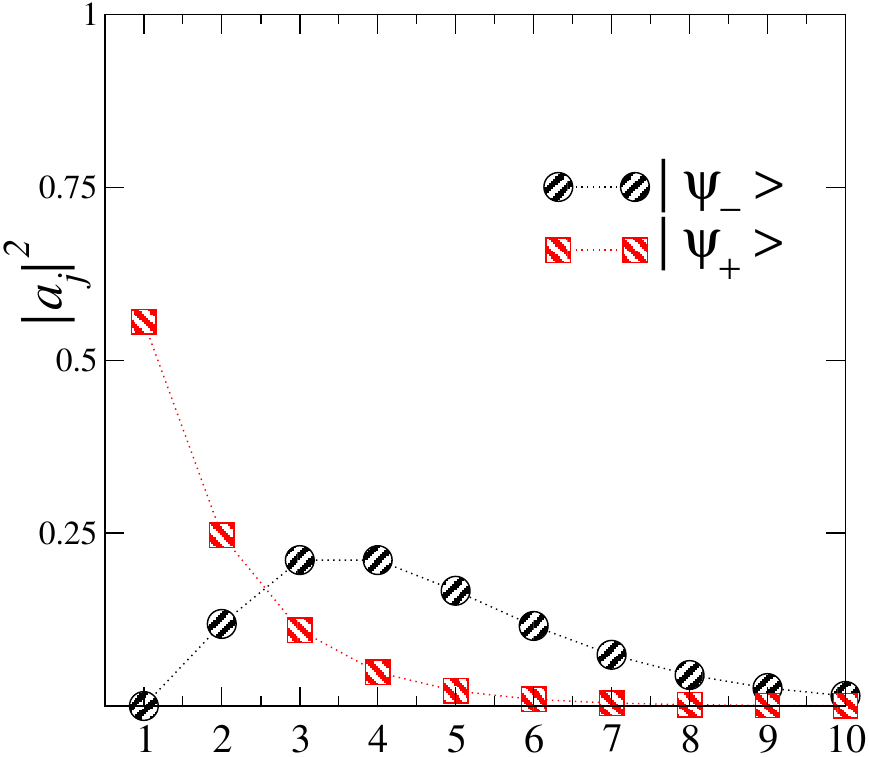}
\includegraphics[scale=.28]{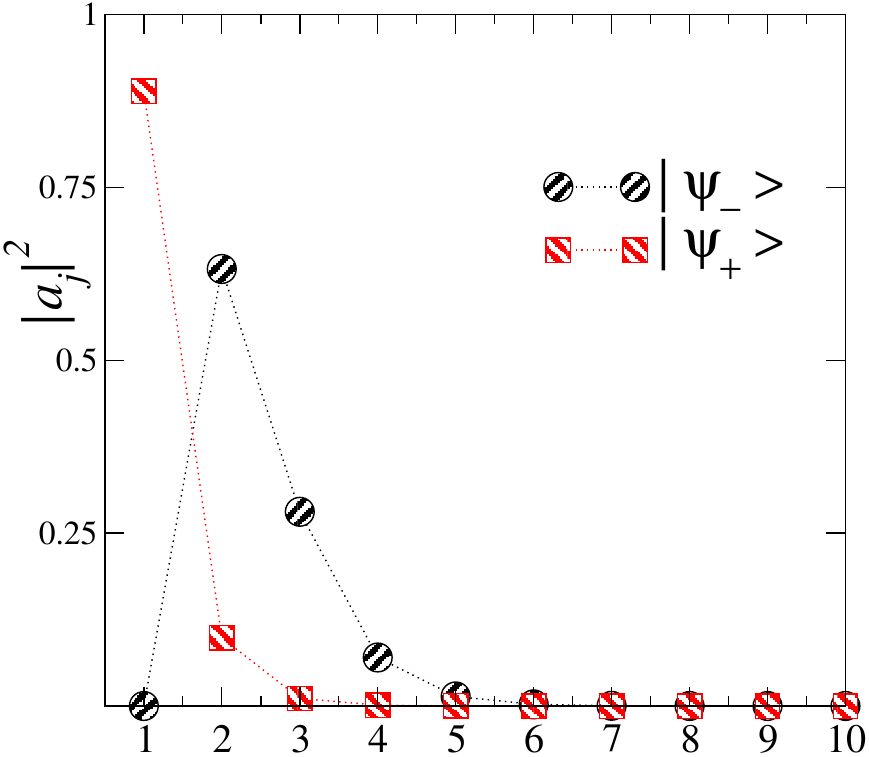}
\end{center}
\caption{(Color online) Weights of the component of the wave functions for two cases. Left: $\eta=1.5$ and Right: $\eta=3$}
\label{sli}
\end{figure}

\section{Summary and discussion}\label{section-summary}

We have employed analytical methods to study the extended Su–Schrieffer–Heeger model, which includes hopping to the $j$-neighbour between atoms that are not equivalent under translational symmetry, with real amplitudes $u_j$ with $j\in\{0,\dots,M\}$.

We show that the resulting number of edge states is consistent with the bulk–boundary correspondence of topological phases. The off-diagonal structure of the Bloch Hamiltonian under periodic boundary conditions [Eq.~(\ref{chiral-bloch-ham})] allows one to compute the winding number,
which takes integer values between 0 and $M$.

To obtain analytical expressions for the edge states of a chain with open boundary conditions, we have restricted our study to the case $M=2$.
Nevertheless, for small values of $M$,
the procedure based on the method of Alase \textit{et al.}  \cite{alase-1,alase-2,aligia-2} can be straightforwardly extended by solving a polynomial equation of degree $M$.
This approach entails a significantly lower computational cost than standard numerical methods, particularly for long chains.

We have derived exact analytical expressions for the edge states of a chain
with open boundary conditions. Moreover, for finite chains we derive approximate analytical expressions for the low-energy edge states and their energies, which show excellent agreement with the numerical results of Ref.~\cite{li}. We also demonstrate that the approximation is highly accurate.

In Ref.~\cite{li2}, the authors tune the experimental setup along a one-parameter trajectory that connects a bulk sector with winding number two (a topologically nontrivial phase) to one with winding number zero (a trivial phase). We derive the corresponding edge states and analyze their structure along this path. Our results reveal nontrivial contributions from the different lattice sites to the spatial weight distribution of these states.

Our findings clarify the fundamental nature of the end states emerging in the extended Su–Schrieffer–Heeger model.

\section*{Acknowledgments}
We are partially supported by CONICET, Argentina.

\appendix
\section{Orthonormalized edge states for winding number 2}\label{appendix-theta0}

In this appendix, we obtain a pair of orthonormalized edge states corresponding to a winding number \(\delta = 2\).
In particular, we show that these states are still present along the line \(1 - 4\tilde{u}_{0}\tilde{u}_{2} = 0\) in Eq.~(\ref{bulk-edge-problem-7}).

In general, $z$ is a complex number, $z = |z| e^{i\phi}$.
If $z$, with $|z| < 1$, is a root of $Q_1(z)$ of Eq.~(\ref{bulk-edge-problem-6}), where $\tilde{u}_2$ is sufficiently large and positive, then $\overline{z}= |z| e^{-i\phi}$ is also a root.
Therefore, we have the following two normalized edge states:
\begin{eqnarray}\label{1-2-states}
|\psi_{A}(z)\rangle &=&\sqrt{1-|z|^2}\sum_{j=1}^{\infty} z^{j-1}|A_j\rangle\nonumber\\
             &=& \sqrt{1-|z|^2}\sum_{j=1}^{\infty}|z|^{j-1}e^{i(j-1)\phi}|A_j\rangle\nonumber\\
|\overline{\psi}_{A}(z)\rangle &=&\sqrt{1-|z|^2}\sum_{j=1}^{\infty} \overline{z}^{j-1}|A_j\rangle\nonumber\\
             &=& \sqrt{1-|z|^2}\sum_{j=1}^{\infty}|z|^{j-1}e^{-i(j-1)\phi}|A_j\rangle.
\end{eqnarray}
The norm is computed using the complex geometric series $\sum_{j=p}^{\infty} z^j = \frac{z^p}{1-z}$, valid for $|z| < 1$.
The overlap of these states is the complex number
\begin{eqnarray}\label{1-2-overlap}
\langle\overline{\psi}_{A}(z)|\psi_{A}(z)\rangle
&=&\frac{(1-|z|^2)(1-|z|^2\cos(2\phi)}{1-2|z|^2\cos(2\phi)+|z|^4}\nonumber\\
&&+i~\frac{(1-|z|^2)(|z|^2\sin(2\phi)}{1-2|z|^2\cos(2\phi)+|z|^4},
\end{eqnarray}
with modulus
\begin{eqnarray}\label{1-2-overlap-modulo}
|\langle\overline{\psi}_{A}(z)|\psi_{A}(z)\rangle|^{2}|
=\frac{1-2|z|^2+|z|^4}{1-2|z|^2\cos(2\phi)+|z|^4}.
\end{eqnarray}

Note that over the curve \(1 - 4\tilde{u}_{0}\tilde{u}_{2} = 0\), $z=\overline{z}$ and therefore $|\psi_{A}(z)\rangle=|\overline{\psi}_{A}(z)\rangle$ suggesting the presence of only one edge state. However, as we will see in what follows, there are two well defined edge states along that curve.
Now we introduce the sum and difference of $|\psi_{A}(z)\rangle$ and $|\overline{\psi}_{A}(z)\rangle$:
\begin{eqnarray}\label{s-d-states}
|\psi_{+}(z)\rangle &=& |\psi_{A}(z)\rangle+|\overline{\psi}_{A}(z)\rangle\\
&=&2\sqrt{1-|z|^2}\sum_{j=1}^{\infty}|z|^{j-1}\cos((j-1)\phi)|A_j\rangle, \nonumber\\
|\psi_{-}(z)\rangle &=& \frac{|\psi_{A}(z)\rangle-|\overline{\psi}_{A}(z)\rangle}{i}\nonumber\\
&=&2\sqrt{1-|z|^2}\sum_{j=1}^{\infty}|z|^{j-1}\sin((j-1)\phi)|A_j\rangle\nonumber.
\end{eqnarray}

Their overlap is given by
\begin{eqnarray}\label{s-d-overlap}
\langle \psi_{-}(z)|\psi_{+}(z)\rangle
&=&\frac{2|z|^2(1-|z|^{2})\sin(2\phi)}{1-2|z|^2\cos(2\phi)+|z|^4}.
\end{eqnarray}
The leading term in the series expansion is of order $O(|z|^2)$. Regarding their inner products, they are:
\begin{eqnarray}\label{s-s-overlap}
\langle \psi_{+}(z)|\psi_{+}(z)\rangle
&=& 2-2\frac{(1-|z|^2)(1-|z|^2\cos(2\phi))}{1-2|z|^2\cos(2\phi)+|z|^4}\nonumber.
\end{eqnarray}

With this, we define the following normalized difference state:
\begin{eqnarray}\label{d1}
|\widehat{\psi_{-}}(z)\rangle &=& \frac{|\psi_{-}(z)\rangle}{\sqrt{\langle \psi_{-}(z)|\psi_{-}(z)\rangle}} \nonumber\\
&=& c_{-}(z)\sum_{j=1}^{\infty}|z|^{j-1}\sin(j\phi)|A_{j+1}\rangle,
\end{eqnarray}
with
\begin{eqnarray}\label{d2}
c^2_{-}(z)= \frac{2(1-|z|^2)(1-2|z|^2\cos(2\phi)+|z|^4)}{(1+|z|^2)(1-\cos(2\phi))}.
\end{eqnarray}

Note that
\[
\frac{\sin(j\phi)}{\sqrt{1-\cos(2\phi)}}
\longrightarrow \mathrm{sgn}(\phi)\, \frac{j}{\sqrt{2}},
\]
so in the limit $\phi \to 0$ we may choose
\begin{eqnarray}\label{d3}
|\widehat{\psi_{-}}(z)\rangle\big|_{\phi=0} &=&
\sqrt{\frac{(1-|z|^2)^3}{1+|z|^2}}
\sum_{j=1}^{\infty} j |z|^{j-1}|A_{j+1}\rangle.
\end{eqnarray}

Similarly, the normalized sum state is:
\begin{eqnarray}\label{s1}
|\widehat{\psi_{+}}(z)\rangle &=& c_{+}(z)
\sum_{j=1}^{\infty}|z|^{j-1}\cos((j-1)\phi)|A_j\rangle,
\end{eqnarray}
where
\begin{eqnarray}\label{s2}
 c_{+}(z)=\sqrt{\frac{2(1-|z|^2)}{
1+\frac{(1-|z|^2)(1-|z|^2\cos(2\phi))}{1-2|z|^2\cos(2\phi)+|z|^4}}}.
\end{eqnarray}

Therefore,
\begin{eqnarray}\label{s3}
|\widehat{\psi_{+}}(z)\rangle\big|_{\phi=0} &=&
\sqrt{1-|z|^2}
\sum_{j=1}^{\infty}|z|^{j-1}|A_{j}\rangle.
\end{eqnarray}
Note that, however, we can choose an orthogonal state to $|\widehat{\psi_{-}}(z)\rangle$ by using the Gram-Schmidt orthogonalization procedure:
\begin{eqnarray}\label{s-prime}
|\psi'(z)\rangle&=&|\widehat{\psi_{+}}(z)\rangle-\langle \widehat{\psi_{-}}(z)|\widehat{\psi_{+}}(z)\rangle|\widehat{\psi_{-}}(z)\rangle,
\end{eqnarray}
where
\begin{eqnarray}\label{s-prime-norm}
 \langle \psi'(z)|\psi'(z)\rangle=1-\lvert\langle \widehat{\psi_{-}}(z)|\widehat{\psi_{+}}(z)\rangle\rvert^2.
\end{eqnarray}
Explicitly:
\begin{eqnarray}\label{s-prime2}
|\widehat{\psi'}(z)\rangle&=&c_{+}(z)\sum_{j=1}^{\infty}|z|^{j-1}\cos((j-1)\phi)|A_{j}\rangle\nonumber\\
&&-a(z)\sum_{j=1}^{\infty}|z|^{j-1}\sin(j\phi)|A_{j+1}\rangle,
\end{eqnarray}
with
\begin{eqnarray}\label{s-prime3}
 a(z)&=&c_{+}(z)c^2_{-}(z)\frac{2(1-|z|^2)|z|^2\sin(2\phi))}{1+|z|^4-2|z|^2\cos(2\phi))}.
\end{eqnarray}

In particular, over the line $u_{1}^{2}=4u_{0}u_{2}$ ($\phi=0$), $a(z)=0$ and we obtain:
\begin{eqnarray}\label{s-prime4}
|\widehat{\psi'}(z)\rangle\big|_{\phi=0}=|\widehat{\psi_{+}}(z)\rangle\big|_{\phi=0}.
\end{eqnarray}

Therefore, along that curve there are two orthonormalized edge states, $|\widehat{\psi_{+}}(z)\rangle\big|_{\phi=0}$ and $|\widehat{\psi_{-}}(z)\rangle\big|_{\phi=0}$ if $|z|<1$ or none otherwise.

\section{Edge states for a finite chain}\label{appendix-finite-chain}
For a chain of $N$ unit cells, the Hamiltonian reads
\begin{eqnarray}\label{H-n}
	H_{N}&=&\sum_{j=1}^{N-2} \Big( u_0 | B_j\rangle\langle A_j|
	+ u_1 | B_j\rangle\langle A_{j+1}|
	+ u_2 | B_j\rangle\langle A_{j+2}| \Big) \nonumber\\
	&&+u_0\Big(| B_{N-1}\rangle\langle A_{N-1}| + | B_{N}\rangle\langle A_{N}|\Big) \nonumber\\
	&&+u_1 | B_{N-1}\rangle\langle A_{N}| + \mathrm{H.c.}
\end{eqnarray}

For simplicity we explain the formalism for the case with winding number 2 and
two complex conjugate states. Extension to other cases is straightforward.
The normalized states analogous to those in Eq.~(\ref{1-2-states}) are
\begin{equation}\label{1-2-N-A-states}
	\begin{array}{l}
		|\psi_{A}(z)\rangle =
		\sqrt{\dfrac{1-|z|^2}{1-|z|^{2N}}}
		\sum_{j=1}^{N} z^{\,j-1}|A_j\rangle,\\[8pt]
		|\overline{\psi}_{A}(z)\rangle =
		\sqrt{\dfrac{1-|z|^2}{1-|z|^{2N}}}
		\sum_{j=1}^{N} \overline{z}^{\,j-1}|A_j\rangle.
	\end{array}
\end{equation}

Similar states, whose maximum weight is localized at the right end of the chain (a $B$ site), are given by
\begin{equation}\label{1-2-N-B-states}
	\begin{array}{l}
		|\psi_{B}(z)\rangle =
		\sqrt{\dfrac{1-|z|^2}{1-|z|^{2N}}}
		\sum_{j=1}^{N} z^{\,j-1}|B_{N-j+1}\rangle,\\[8pt]
		|\overline{\psi}_{B}(z)\rangle =
		\sqrt{\dfrac{1-|z|^2}{1-|z|^{2N}}}
		\sum_{j=1}^{N} \overline{z}^{\,j-1}|B_{N-j+1}\rangle.
	\end{array}
\end{equation}

Note that if $|z|<1$, then
\[
\sqrt{\frac{1-|z|^2}{1-|z|^{2N}}} \simeq \sqrt{1-|z|^2}
\]
for sufficiently large $N$.

Acting with the Hamiltonian on $|\psi_{A}(z)\rangle$, the contribution from the sum
$\sum_{j=1}^{N-2}$ in Eq.~(\ref{H-n}) produces a term proportional to $Q_1(z)$ and therefore vanishes.
The remaining terms yield
\begin{eqnarray}\label{H-sobre-1a}
	H_{N}|\psi_{A}(z)\rangle
	&=&
	\sqrt{1-|z|^2}
	\Big(z^{N-1}u_0| B_N\rangle
	-z^{N}u_2| B_{N-1}\rangle\Big)\nonumber.
\end{eqnarray}

Similarly, for $|\overline{\psi}_{A}(z)\rangle$, since $Q_1(\overline{z})=0$, we obtain
\begin{eqnarray}\label{H-sobre-2a}
	H_{N}|\overline{\psi}_{A}(z)\rangle
	&=&
	\sqrt{1-|z|^2}
	\Big(\overline{z}^{N-1}u_0| B_N\rangle
	-\overline{z}^{N}u_2| B_{N-1}\rangle\Big)\nonumber.
\end{eqnarray}

As a result, the matrix elements are
\begin{equation}\label{elementos-H}
	\begin{array}{l}
		\langle \psi_{B}(z)| H_{N} |\psi_{A}(z)\rangle = \alpha_1,\\[6pt]
		\langle \psi_{B}(z)| H_{N} |\overline{\psi}_{A}(z)\rangle = \overline{\alpha}_2,\\[6pt]
		\langle \overline{\psi}_{B}(z)| H_{N} |\psi_{A}(z)\rangle = \alpha_2,\\[6pt]
		\langle \overline{\psi}_{B}(z)| H_{N} |\overline{\psi}_{A}(z)\rangle = \overline{\alpha}_1,
	\end{array}
\end{equation}
with $\alpha_1 = z^{N-1}(1-|z|^2)\big(u_0-|z|^2u_2\big)$, and $\alpha_2 = z^{N-1}(1-|z|^2)\big(u_0-z^{2}u_2\big)$.
The set
\[
\mathcal{H}_{\mathrm{low}}
=\big\{
|\psi_{A}(z)\rangle,
|\overline{\psi}_{A}(z)\rangle,
|\psi_{B}(z)\rangle,
|\overline{\psi}_{B}(z)\rangle
\big\}
\]
defines a Hilbert subspace that is sufficient to describe, approximately yet accurately, the low-energy states localized at the edges. Following an argument similar to that of Ref.~\cite{aligia-2}, one can show that the effect of the remaining states in the full Hilbert space on the lowest-energy levels is of order $|z|^{2N}/\Delta_g$.

The matrix elements of the restriction of $H_{N}$ to this subspace form the following matrix:
\begin{eqnarray}\label{elementos-H-1}
	\left[ H_{N}\big|_{\mathcal{H}_{\mathrm{low}}}\right]
	=\left(\begin{array}{cccc}
		0 & 0 & \overline{\alpha}_1 & \overline{\alpha}_2\\
		0 & 0 & \alpha_2 & \alpha_1\\
		\alpha_1 & \overline{\alpha}_2 & 0 & 0\\
		\alpha_2 & \overline{\alpha}_1 & 0 & 0
	\end{array}\right).
\end{eqnarray}

Our goal is to solve the Schr\"odinger equation within this subspace,
\[
H_{N}\big|_{\mathcal{H}_{\mathrm{low}}}|\psi\rangle=E|\psi\rangle,
\]
where
\[
|\psi\rangle
=a_1|\psi_{A}(z)\rangle
+a_2|\overline{\psi}_{A}(z)\rangle
+b_1|\psi_{B}(z)\rangle
+b_2|\overline{\psi}_{B}(z)\rangle.
\]
Since the basis is non-orthogonal, the Schr\"odinger equation leads to the generalized eigenvalue problem
\begin{eqnarray}\label{sistema-1}
	\left[ H_{N}\big|_{\mathcal{H}_{\mathrm{low}}}\right]
	\left(\begin{array}{c}
		a_1 \\ a_2 \\ b_1 \\ b_2
	\end{array}\right)
	=
	E S
	\left(\begin{array}{c}
		a_1 \\ a_2 \\ b_1 \\ b_2
	\end{array}\right),
\end{eqnarray}
where the overlap matrix is
\begin{eqnarray}\label{overlap-matrix}
	S= \left(\begin{array}{cccc}
		1 & s & 0 & 0\\
		\overline{s} & 1 & 0 & 0\\
		0 & 0 & 1 & s\\
		0 & 0 & \overline{s} & 1
	\end{array}\right),
\end{eqnarray}
with
$s=\langle\psi_{A}(z)|\overline{\psi}_{A}(z)\rangle
=\langle\psi_{B}(z)|\overline{\psi}_{B}(z)\rangle$.

Equation~(\ref{sistema-1}) can be rewritten as
\begin{eqnarray}\label{sistema-2}
	\left(\begin{array}{c}
		X b \\ X^\dagger a
	\end{array}\right)
	=
	E\left(\begin{array}{c}
		Y a \\ Y b
	\end{array}\right),
\end{eqnarray}
where
\begin{eqnarray}\label{sistema-3}
	X&=&\left(\begin{array}{cc}
		\overline{\alpha}_1 & \overline{\alpha}_2 \\
		\alpha_2 & \alpha_1
	\end{array}\right),\qquad
	Y=\left(\begin{array}{cc}
		1 & s \\
		\overline{s} & 1
	\end{array}\right),\nonumber\\
	a&=&\left(\begin{array}{c}a_1\\a_2\end{array}\right),\qquad
	b=\left(\begin{array}{c}b_1\\b_2\end{array}\right).
\end{eqnarray}

Since $Y$ is nonsingular, it follows that
\begin{eqnarray}\label{sistema-4}
	Y^{-1}X^{\dagger}a&=&Eb,\nonumber\\
	Y^{-1}Xb&=&Ea.
\end{eqnarray}
Because $X=X^{\dagger}$, both vectors $a$ and $b$ satisfy the same equation,
\begin{equation}\label{sistema-5}
	h^2 c = E^2 c,
\end{equation}
where $c=(c_1,c_2)^T$ and $h=Y^{-1}X$.
Nontrivial solutions require $\det(h^2-E^2)=0$, which determines the allowed values of $E^2$.

We now focus on the case where $z$ is a complex solution of $Q_1(z)=0$.
In this situation, $\alpha_1=0$, since $u_0-|z|^2u_2=0$, independently of $u_1$.
Thus,
\[
X=\left(\begin{array}{cc}
	0 & \overline{\alpha}_2\\
	\alpha_2 & 0
\end{array}\right),
\qquad
Y^{-1}=\frac{1}{1-|s|^2}
\left(\begin{array}{cc}
	1 & -s\\
	-\overline{s} & 1
\end{array}\right).
\]
This yields
\begin{eqnarray}\label{energy-1}
	E^2_\pm
	&=&
	\frac{|\alpha_2|^2+\mathrm{Re}(\alpha_2^2s^2)}{(1-|s|^2)^2}
	\\
	&&\pm
	\frac{\sqrt{\big(|\alpha_2|^2+\mathrm{Re}(\alpha_2^2s^2)\big)^2
			+(1-|s|^2)^2|\alpha_2|^4}}{(1-|s|^2)^2}\nonumber.
\end{eqnarray}
Note that since $\alpha_2$ is of order $z^{N-1}$, the values of $|E|$ scale as $|z|^{(N-1)}$, while the estimated error $\sim |z|^{2N}/\Delta_g$ decays much faster. Consequently, $\mathcal{H}_{\mathrm{low}}$ constitutes a good approximation
for the low-energy sector for a sufficiently long chain.

In summary, Eqs.~(\ref{sistema-4}), (\ref{sistema-5}), and (\ref{energy-1}) yield two pairs of edge states $|\psi\rangle$.


\begin{thebibliography}{99}
\bibitem{note2} Note that $P_bP_e=P_eP_b=0$ and $P^2_b=P^2_e=1$. From $P_b+P_e=1$, one obtains $P_bH|\psi\rangle + P_eH|\psi\rangle = EP_b|\psi\rangle + EP_e|\psi\rangle$. Applying $P_b$ to this equation, we obtain  $P_b H |\psi\rangle = EP_b  |\psi\rangle$. Similarly, applying $P_e$ leads to $P_e H |\psi\rangle = EP_e  |\psi\rangle$.
\bibitem{note} Note that in a finite chain, the edge states appear in pairs, with each one localized at one end of the chain. For the states localized at the right end of the chain $|z_i| > 1$.
\end{thebibliography}
\end{document}